# Noncontact Imaging of Ion Dynamics in Polymer Electrolytes with Time-Resolved Electrostatic Force Microscopy

*Jeffrey S. Harrison[1], Dean A. Waldow[2], Phillip A. Cox[1], Rajiv Giridharagopal[1], Marisa L. Adams[2], Victoria L. Richmond[2], Sevryn P. Modahl[2], Megan R. Longstaff[2], Rodion A. Zhuravlev[2], David S. Ginger[1*]*

[1]Department of Chemistry, University of Washington, Seattle, Washington 98195, United States

[2]Department of Chemistry, Pacific Lutheran University, Tacoma, Washington 98447, United States

*Address correspondence to ginger@chem.washington.edu

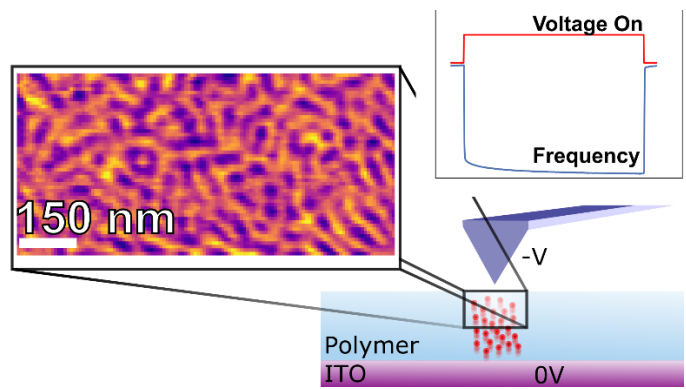

**Table of contents graphic**

## *Abstract*

Ionic transport processes govern performance in many classic and emerging devices, ranging from battery storage to modern mixed-conduction electrochemical transistors. Here, we study local ion transport dynamics in polymer films using time-resolved electrostatic force microscopy (trEFM). We establish a correspondence between local and macroscopic measurements using local trEFM and macroscopic electrical impedance spectroscopy (EIS). We use polymer films doped with lithium bis(trifluoromethane)sulfonimide (LiTFSI) as a model system where the polymer backbone has oxanorbornenedicarboximide repeat units with an oligomeric ethylene oxide side chain of length n. Our results show that the local polymer response measured in the time domain with trEFM follows stretched exponential relaxation kinetics, consistent with the Havriliak-Negami relaxation we measure in the frequency-domain EIS data for macroscopic samples of the same polymers. Furthermore, we show that the trEFM results capture the same trends as the EIS results — changes in ion dynamics with increasing temperature, increasing salt concentration, and increasing volume fraction of ethylene oxide side chains in the polymer matrix evolve with the same trends in both measurement techniques. We conclude from this correlation that trEFM data reflect, at the nanoscale, the same ionic processes probed in conventional EIS at the device level. Finally, as an example application for emerging materials syntheses, we use trEFM and infrared photoinduced force microscopy (PiFM) to image a novel diblock copolymer electrolyte for next-generation solid-state energy storage applications.

Ion transport processes in thin film materials affect the performance of a wide variety of technologies ranging from bioelectronics[1,2] and electrochemical energy storage[3], to the stability of perovskite photovoltaics.[4,5] The ability to characterize ion transport dynamics in solid-state systems at the nanoscale is important to understanding the processing-structure-function relationships of these materials and has therefore attracted considerable interest.[6–8] Many polymer-based systems are, either by design or by consequence of their solution-based deposition kinetics, inherently nanostructured.[9,10] Consequently, to explore the properties of such materials, groups have used a wide range of scanning probe microscopy methods,[11–14] as well as measuring the behavior of a moving ion front via optical methods.[2,8,15] Electrochemical strain microscopy (ESM) in particular has been widely applied to image ion uptake dynamics[16,17] and our group reported the use of ESM to probe local swelling behavior upon ion uptake in polymer electrochemical transistors.[18,19]

Nevertheless, methods like ESM were originally developed for high-modulus inorganic materials and can be challenging to apply on soft polymer and gel samples. Non-contact methods offer many advantages for imaging dynamics of soft samples, and methods like time-resolved electrostatic force microscopy have been widely employed to image dynamic processes in organic electronic materials.[20–23]

Several groups have used similar time-domain frequency shift measurements to record local ionic responses in inorganic materials following voltage pulses, but with limited imaging.[24–27] Here, we explore the use of trEFM to make local measurements of ion dynamics in soft polymer systems, including imaging, and we compare nanoscale trEFM measurements against macroscopic electrochemical impendence measurements to verify the correspondence of these methods under a wider range of conditions.

As a model system, we use a polymer electrolyte material synthesized for use in energy storage applications – a polymer with an oxanorbornenedicarboximide backbone repeat unit and an oligomeric ethylene oxide side chain of length n (polyONDI-n) (**Fig 1a**) where n indicates the number of repeating ethylene oxide sidechain units. The synthesis of this material has been previously reported and the bulk ionic conductivity has been extensively characterized.[28] This polymer electrolyte is an ideal model system for testing the correlation between trEFM and EIS as we can control ion transport and relaxation properties through a variety of methods including temperature, dopant ion concentration, and polymer side-chain length.

First, we compare microscopic kinetic data obtained from trEFM on poly(ONDI-n) homopolymer films with conventional EIS measurements on the same films. Time-resolved electrostatic force microscopy (trEFM) is a non-contact method for tracking local changes in electrical properties of a sample over time.[29] **Fig. 1a** depicts the schematic approach we use to apply trEFM to study ion motion in a soft polymer. In a typical measurement cycle, we apply a - 10 V bias to the cantilever of an atomic force microscope (AFM). This bias causes positive ions within the sample to migrate towards the tip (negative ions away from the tip), causing a decrease in the resonant frequency of the cantilever system. We track this frequency as a function of time to extract the relevant dynamic information.

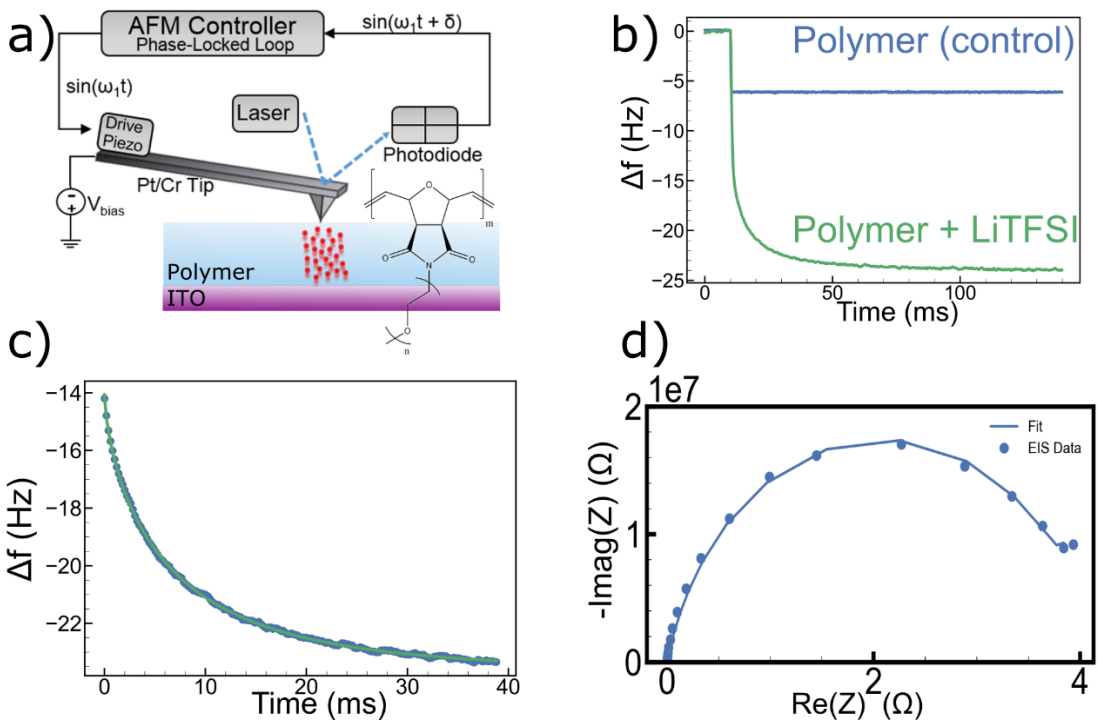

**FIGURE 1. Comparison of time domain (microscopic) and frequency domain (macroscopic) measurements of ionic relaxation on poly(ONDI-4) films doped with 10% (wt/wt) LiTFSI salt at 25 °C** (a) Schematic representation of the trEFM process. (b) A trEFM measurement showing the cantilever frequency shift (Δf) as a function of time after a -10V voltage pulse is turned on above an undoped polymer (control, blue)) and polymer doped with LiTFSI (green). The polymer with dissolved LiTFSI salt shows ion relaxation dynamics that span microsecond to millisecond times. (c) The trEFM frequency shift as a function of time, zoomed in at the onset of ionic relaxation. Open circles are experimental data, the green solid line is a fit to the stretched exponential function with $\tau = 32\ ms$ and $\beta = 0.82$ (d) Cole-Cole Plot of electrochemical impedance spectroscopy (EIS) data taken from $10^5$ to $10^{-2}$ (Hz). The solid circles are the experimental data points, and the solid blue line is a fit to the Havriliak-Negami (Eq. 1) with α=0.89, β=0.94, R = 39 MΩ, $\langle \tau_{EIS} \rangle = 106\ ms$, see main text for discussion.

More precisely, the redistribution of the mobile ions in response to the applied bias screens the electrostatic force between the tip and substrate, causing a change in the local tip-sample electrostatic force gradient, which alters the resonant frequency of the cantilever. By recording the resonant frequency shift vs. time using a frequency-shift feedback loop, we can thus use trEFM to record the dynamic relaxation of the local ions. The change in resonance frequency that results from a change in the electrostatic force gradient[30] between a nanoscale tip and sample, Δf, is given by $\Delta f = \frac{-f_0}{2k}\frac{\partial^2 C}{\partial z^2}V^2$, where $f_0$ is the natural frequency of the cantilever, C is the tip-sample capacitance, z is the mean distance above the sample, and V is the electric potential difference between the cantilever and grounded substrate. The time evolution of this frequency shift therefore records the ion motion under the cantilever following the voltage pulse.

**Fig. 1b** shows such a trEFM frequency vs. time trace following application of a -10V bias applied to the EFM cantilever at an arbitrary point above an undoped (ion-free) poly(ONDI-4) polymer sample (solid blue line) and another trace for a film of the same polymer doped with 10% (wt/wt) lithium bis(trifluoromethane)sulfonimide (LiTFSI) (solid green line). The bias pulse was turned on at time t=0. **Fig. 1b** shows the cantilever responds rapidly above the pure (undoped) polymer: the trace resembles a step function with a rise time on the ~microsecond timescale, comparable to the instrument response function at the gain settings we use herein (see SI Fig. 1). This behavior is consistent with the rapid change in resonant frequency of the cantilever as the potential difference between the tip and sample changes, with minimal slow dielectric relaxation. In contrast, the doped polymer sample shown in **Fig. 1b** behaves much differently, exhibiting a long, slow, decay lasting tens of milliseconds, and reaching a larger frequency shift, Δf, of 25 Hz. We interpret the difference between the salt-doped and undoped polymers as resulting from slow dielectric relaxation due to ion motion in the polymer.

We fit this slow relaxation as seen in **Fig. 1c** using a stretched exponential function, as is commonly observed in both frequency and time-domain studies of ion motion and dielectric relaxation in polymers.[31,32] The form of the frequency shift, Δf, is a stretched exponential given by $\Delta f = \Delta f_{ion} \exp\left(\left(t/\tau_{fit}\right)^{\beta}\right) + \Delta f_0$, where $\Delta f_{ion}$ is the frequency shift due to slow processes such as ionic relaxation, $\Delta f_0$ is the frequency shift occurring on fast times (before ionic relaxation occurs), $\tau_{fit}$ is the characteristic time-constant, and β is the stretching coefficient. Interpreting the stretched exponential as a distribution of single-exponentials, we define the average relaxation time constant of this distribution $\langle \tau \rangle = \frac{\tau_{fit}}{\beta} \Gamma\left(\frac{1}{\beta}\right)$:[33] where $\tau_{fit}$ is the time-constant from the stretched exponential function, β is the stretching coefficient, and Γ is the gamma function.[34] Using this average relaxation, $\langle \tau \rangle$, we can then define the trEFM relaxation rate, $k_{trEFM}$, as the inverse of this average relaxation time constant, $\frac{1}{\langle \tau \rangle}$.

To better understand how these local stretched-exponential relaxation curves compare to more conventional macroscopic measurements of dielectric relaxation in ion-doped polymers, we next perform classical electrical impedance spectroscopy[35] and trEFM on samples prepared simultaneously (See Methods for sample preparation and details of the EIS experiments). **Fig. 1d** shows a typical EIS spectrum obtained on a poly(ONDI-2) sample doped with 10% (wt/wt) LiTFSI, taken at 25 °C.

We fit the electrochemical impedance data in **Fig. 1d** using the well-known Havriliak-Negami equation:[36]

$$Z_{HN} = \frac{R}{(1 + (i\omega \langle \tau_{EIS} \rangle)^{\alpha})^{\beta}} \quad (1)$$

where $\alpha, \beta$ are coefficients related to the distribution of underlying relaxation time-constants, $\langle \tau_{EIS} \rangle$ is the average time-constant of the system, and R is the impedance of the sample in the low-frequency limit. Again, we define the average relaxation rate $k_{EIS}$ as the inverse of the average time-constant, $\frac{1}{\langle \tau_{EIS} \rangle}$. The solid line in **Fig. 1d** is a fit of the experimental EIS data with α=0.89, β=0.94, R = 39 MΩ, $\langle \tau_{EIS} \rangle = 106 ms$ and shows that the Havriliak-Negami function fits the experimental EIS data very well.

**Fig. 1c** shows local trEFM data taken on an identical poly(ONDI-2) film doped with the same concentration of LiTFSI and at the same temperature as the film used for the EIS spectroscopy in **Fig. 1d**. We again fit the trEFM data in **Fig. 1c** using a stretched exponential. Our observation of Havriliak-Negami relaxation dynamics in the frequency domain, and stretched exponential relaxation in the time domain is consistent with these methods measuring the same underlying physical response: the Havriliak-Negami equation can be thought of as a frequency-domain analog of the time-domain stretched exponential function.[37] Both functions often arise from dispersive behavior due to a distribution of characteristic time constants being measured in parallel. We thus take our observation of stretched exponential kinetics via trEFM and Havriliak-Negami relaxation in the frequency domain as evidence that trEFM is indeed measuring the same relaxation processes as EIS.

To further check that the EIS data and trEFM data are measuring the same underlying physical phenomena, we next explore how the relaxation kinetics measured by both methods compare as a function of temperature,[24] dopant concentration, and length of ethylene oxide sidechain on the poly(ONDI-n).

For the first comparison as a function of temperature and ion concentration, we keep the sidechain length constant at n=3. There is strong qualitative agreement between the trends observed for the EIS time constants and the trEFM time constants. We observe that both bulk and nanoscale measured average rates show activated behavior in response to changes in temperature, as well as faster transport with increasing salt concentrations **(Fig 2a, b)**. Both bulk and nanoscale measurements thus show faster relaxation for samples with higher degrees of ionic conductivity.

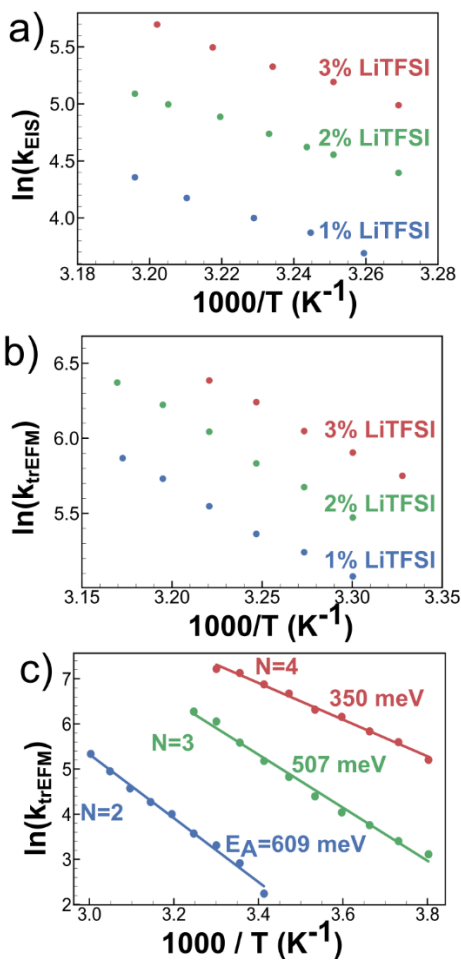

**FIGURE 2. Comparison of trEFM and bulk relaxation rates under temperatures, salt concentrations, and polymer side chain lengths.** (a) Arrhenius plots of the natural logarithm of the relaxation rate determined by EIS plotted against reciprocal T for three different concentrations of LiTFSI salt (3% red, 2% green, 1% blue) (b) The natural logarithm of the relaxation rate as determined by trEFM plotted against reciprocal temperature for three different concentrations of LiTFSI salt. The relaxation rate becomes faster increasing salt concentration and increasing temperature; the same behavior observed in the bulk measurement. (c) Arrhenius analysis of the trEFM relaxation rate reveals that the activation energy of the ionic transport process decreases with increasing sidechain length (n=2 red, n=3 green, n=4 blue).

With the traditional behavior as a function of temperature and dopant concentration established, we next look at the result of altering the molecular structure of the polymer electrolyte. The ion conducting abilities of the polymer electrolyte are commonly understood to result directly from the interaction between the ethylene oxide (EO) sidechain and the dopant ions.[28,38–40] In **Fig. 2c,d** we demonstrate how increasing the length of the EO sidechain leads to an increase in the ability of the electrolyte to transport solvated ions. Three polymer systems with sidechain lengths n= 2, 3, and 4 were used to produce films with similar thicknesses of ~150 nm. The concentration of LiTFSI was held constant so that any change in ion dynamics results directly from the different sidechain identity without incorporating effects due to concentration differences. If we compare the results of the two experiments by plotting the results against each other (**Fig 3.**), we observe a similar positive linear correlation. Adding together the dependence of relaxation rate on temperature, dopant concentration, and sidechain length, we show that trEFM captures information about the same relaxation processes that can be observed in a traditional bulk measurement. Using this result, we assert that a change in the trEFM relaxation rate indicates a change in time-constant caused by variation in ion conductivity in sub-volume probed by the AFM tip. We speculate that the reason the relaxation rates do not have an exact one-to-one correspondence could result from a difference in parallel capacitances due to the vastly different geometries of the device under test in each experimental setup resulting in different sensitivities to different relaxation frequency bands.[41]

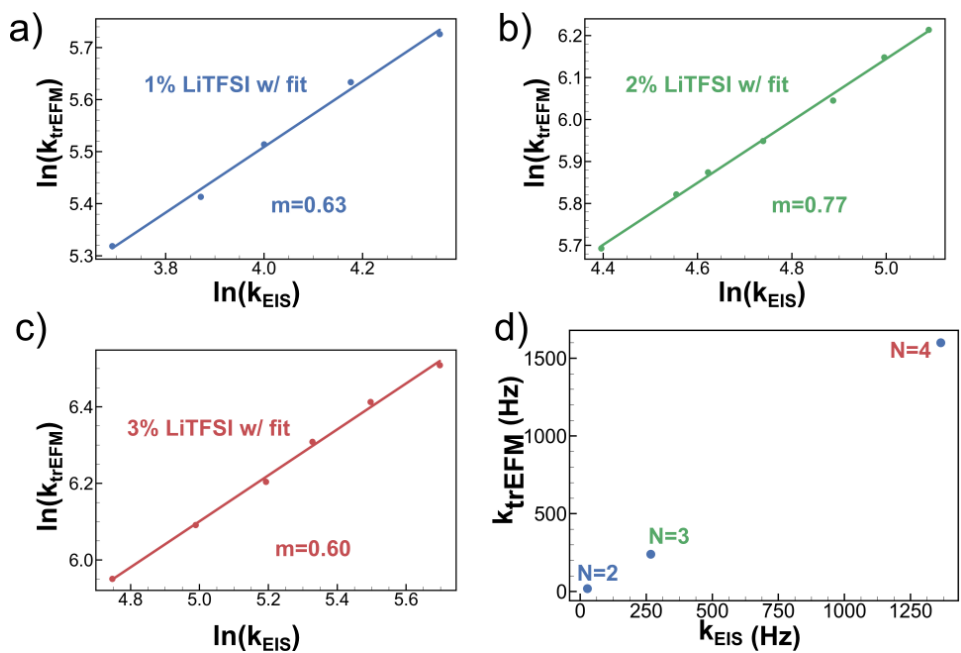

**FIGURE 3. Comparison of bulk EIS to local trEFM.** (a-c) The trEFM relaxation rate is plotted against the EIS relaxation rate, revealing a linear relationship between the two measurements. (d) A comparison of the relaxation rates determined by trEFM and EIS. In both measurements, the relaxation rate increases as a function of increasing sidechain length. The monotonic, positive, correlation indicates that the local trEFM relaxation rates are directly correlated with the average bulk results measured by EIS.

We use trEFM to characterize a material with significant nanostructured contrast in chemical and ion-transport properties: a diblock copolymer electrolyte system that exhibits substantial changes in ionic conductivity with spatial dependence. The diblock copolymer system we characterize here consists of the same backbone repeat unit as the homopolymers above but with a phenyl containing block exhibiting poor ionic conductivity and a block containing ion transporting oligomeric ethylene oxide units (60/40 w/w respectively, n=12). The diblock synthesis is described in detail in the Methods section. This diblock system is a polymer electrolyte that readily phase separates into ionically conductive and non-conductive regions. For this experiment we dissolve LiTFSI together with the diblock polymer in dichloromethane and spin coat it onto an ITO substrate to produce a film with ~10% (wt/wt) LiTFSI. After spin-coating, we used dichloromethane to solvent vapor anneal the sample,[42] to promote phase separation and alignment.

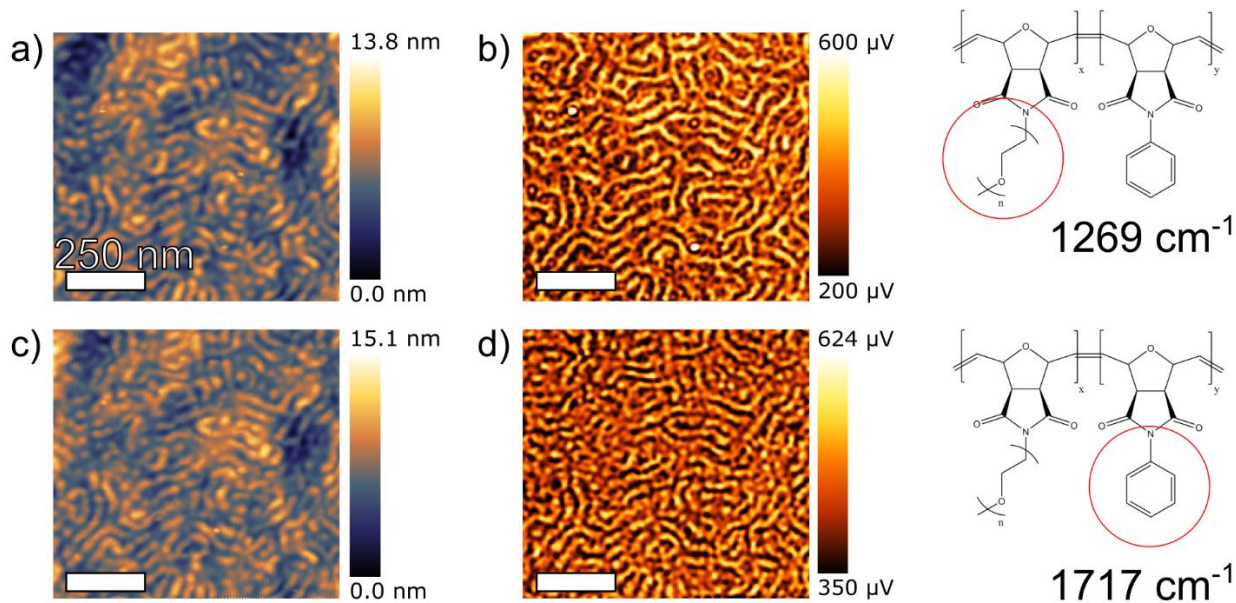

**FIGURE 4. Chemical imaging and identification with PiFM.** (a, b) AFM Topography (a) and corresponding PiFM image taken at 1269 cm$^{-1}$ using a 300 kHz cantilever (b), which corresponds to the ethylene oxide sidechain on the ionically conductive block. (c, d) AFM Topography (c) and corresponding PiFM image taken at 1717 cm$^{-1}$ (d), which corresponds to the phenyl group of the non-conductive block. The phenyl group is found in regions of high topography; where our trEFM data indicates a low charging rate and we would expect slower ion dynamics.

We use photo-induced force microscopy (PiFM)[43–45] to obtain nanoscale chemical composition information to identify the ion transporting and inert regions of the diblock system. We identify two distinct photo-induced force signals, one at an incident excitation at 1269 cm$^{-1}$ which we assign to the ethylene oxide group on the conductive block, and another at 1717 cm$^{-1}$ corresponding to the phenyl group located on the non-conductive block. **Figure 4a** displays the AFM topography, and **Fig. 4b** displays the PiFM signal taken at 1269 cm$^{-1}$. We observe the highest 1269 cm$^{-1}$ signal in areas of relatively low topography, indicating the presence of ion transporting ethylene oxide in these regions. **Fig. 4c** and **Fig. 4d** show the topography and PiFM signal data collected at 1717 cm$^{-1}$, respectively. In contrast to the 1269 cm$^{-1}$ data, the 1717 cm$^{-1}$ scan shows the highest signal in areas of relatively high topography, indicating the presence of the insulating phenyl block of the copolymer.

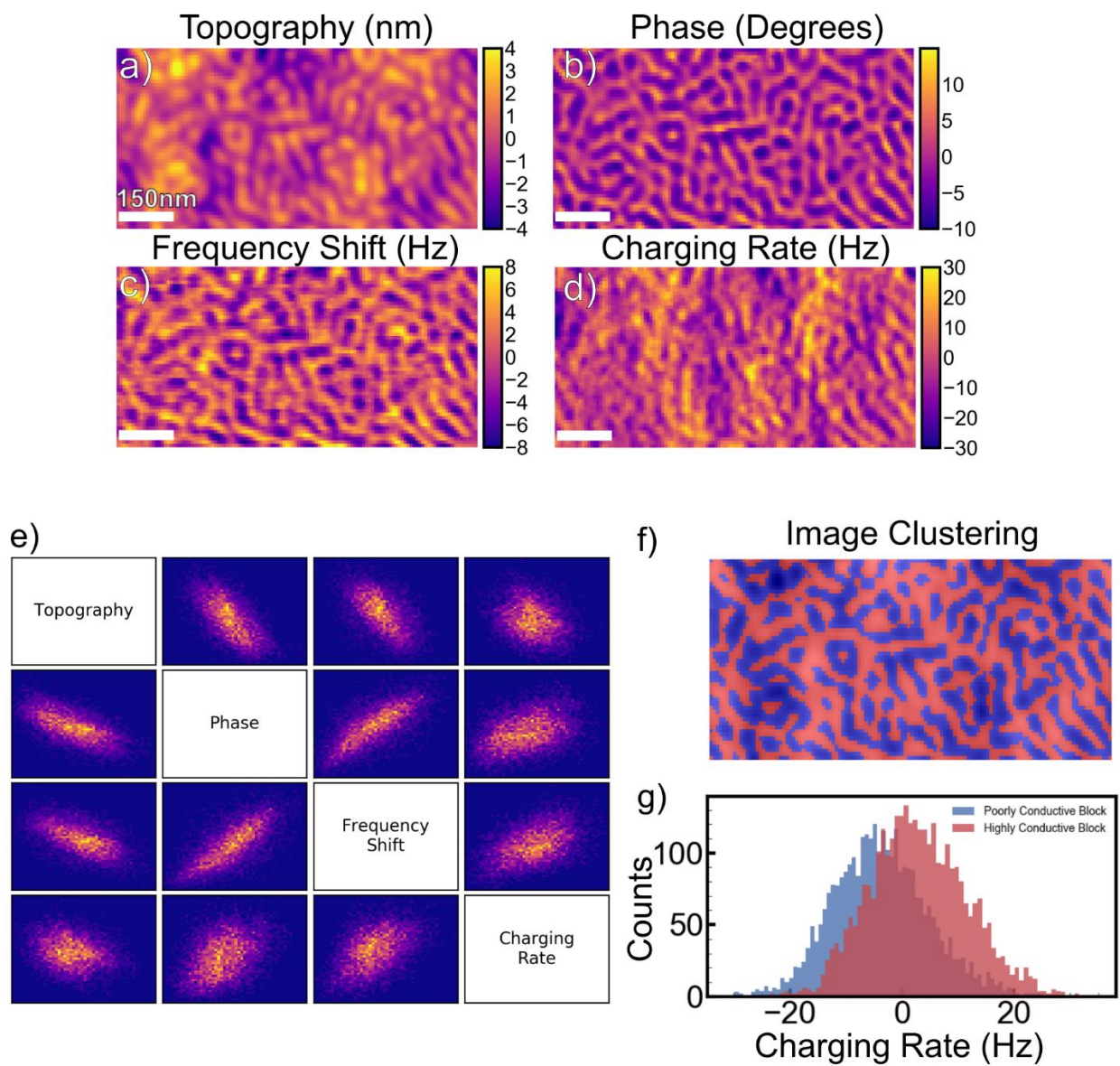

**FIGURE 5. Imaging ionic transport in a diblock copolymer electrolyte.** The value of the data here has been offset around the mean value of the image, showing the relative relationship. (a, b) Intermittent contact

AFM topography and phase images of the diblock polymer. Regions that are high in phase correspond with the ion transporting ethylene oxide block. All data were collected in repulsive mode and as such the phase difference between the cantilever and drive signals are less than 90° (c) The trEFM frequency shift, indicating the areas of the film with the greatest frequency shift correspond with the ion transporting block of the polymer. (d) The trEFM relaxation rate (or charging rate) showing the time-dependent response of the ionic processes within the diblock system. Areas with a higher charging rate indicate the presence of the ion transporting ethylene oxide functionalized block. (e) An analysis showing the correlation between the four images in a-d. The trEFM measurements show high correlation with the phase image, while slightly lower correlation with a topography measurement. (f) The result of an image clustering analysis using Gaussian mixtures of the topography and phase data. Pixels colored in red correspond to the highly ionically conductive region of the film while the pixels colored in blue correspond to the poorly conductive region. (g) The distribution of charging rates within each block normalized to the average charging rate of the entire image. The highly conductive region has a higher average charging rate than that of the poorly conductive block.

Having identified the location of the respective blocks of the copolymer, we next show that the trEFM data agree with PiFM domain assignments. **Fig. 5c and 5d** show the result of the trEFM imaging process. At each pixel of the image the frequency shift as a function of time is collected and averaged together. We fit the time-dependent frequency shift with a stretched exponential function as previously outlined and extract the average relaxation rate. This relaxation rate is fit at each pixel to build an image consisting of trEFM relaxation rates as well as the overall amplitude of the frequency shift. The topography and phase images (**Fig. 5a and 5b**) show two distinct regions with structure on the order of 25 nm. The phase image indicates the presence of a soft region with high phase, and a relatively stiff region with low phase. The soft region corresponds to the ion transporting block of the polymer and is generally located in areas of relatively low topography (which we have previously identified as the ethylene oxide phase by IR PiFM, see **Fig. 4**). The stiff, topographically-raised, region corresponds to the non-conductive phenyl block and results in areas of relatively high topography. The nanoscale variation in ionic relaxation evident in the trEFM frequency shift and charging rate images is clearly correlated with the mechanical and chemical identification that we measured using AFM and PiFM techniques. These images show the presence of a component with a large frequency shift and correspondingly fast relaxation rate which is associated with the presence of the ionically conductive block of the copolymer. Conversely, the areas of the trEFM image with small frequency shift and low relaxation rate are identified with the poorly conductive phenyl block of the copolymer system. To more clearly demonstrate this correlation, we plot the topography, cantilever phase, frequency shift, and charging rate images against each other to produce a correlation matrix (**Fig. 5e**). **Fig. 5** clearly shows that the frequency shift data are anticorrelated with topography and correlated with the cantilever phase.

To better demonstrate that the conductive region of the diblock is associated with a higher trEFM charging rate, we utilize a technique from machine learning known as a Gaussian mixture classification, where the data from **Fig. 5a** and **Fig. 5b** is used as input for the algorithm to categorize each pixel into one of two classes with a unique label. Assigning a color to each label allows the visualization of this classification in **Fig. 5f**. Close examination of this result shows that the previous visual assignments of the regions are also found statistically by assuming the Gaussian distribution of two separate and unique components. **Fig. 5g** shows the distribution of charging rates in these regions normalized to the average charging rate for the entire image; these data show that the conductive regions have a higher average charging rate, while the non-conductive regions have a comparatively lower charging rate. We note that the frequency shift and charging rate distributions for the separate domains are overlap, yet still show distinct character – a result we might expect given the extremely small size (~25 nm) of the domains, and the resolution of EFM methods generally.

In summary, we have studied local ionic relaxation rates measured with trEFM on soft polymer electrolytes, and found that they correspond to the nanoscopic ionic conductivity within the probed sub-volume of the AFM tip. We used EIS measurements to show that there is a close, monotonic correspondence between the changes in trEFM relaxation rate and bulk measurements of the same ionic conductor as a function of temperature, ion concentration, and chemical structure of the host matrix. We conclude that the stretched exponential behavior of the ionic relaxation seen by trEFM and the Havriliak-Negami relaxation observed in the frequency domain are manifestations of the same underlying transport dyanmics, given the stretched-exponentials can be considered time-domain analogs of Havriliak-Negami dyanmics, and both processes can reflect distributed kinetics common in ionic systems. Having further validated trEFM as a viable measure of ion dynamics, we demonstrate the ability to differentiate between conductive and nonconductive regions of a diblock polymer electrolyte. Imaging these materials with a resolution greater than 25 nm allows for immediate application to a wide variety of nanostructured and ionically conductive materials.

## *Acknowledgements*

This paper is based primarily on work supported by the National Science Foundation, NSF DMR-1607242. We gratefully acknowledge graduate fellowship support for J. S. H. from the University of Washington Clean Energy Institute. Part of this work was conducted at the Molecular Analysis Facility, a National Nanotechnology Coordinated Infrastructure site at the University of Washington which is supported in part by the National Science Foundation (grant ECC-1542101), the University of Washington, the Molecular Engineering & Sciences Institute, the Clean Energy Institute, and the National Institutes of Health. D. A. W. acknowledges National Science Foundation awards RUI: DMR-1710549 and MRI-07233226. The authors of this manuscript declare no conflicts of interest or bias that influenced the conclusions expressed within this work.

## *Methods*

*Synthesis*

All materials used as well as the ring-opening metathesis synthesis of the homopolymers are the same as previously published (ref 28) except for the diblock copolymer. The diblock copolymer is synthesized using an oligomeric ethylene oxide side chain with an average length of n=12 for the ion conducting portion and a phenyloxanorbornenedicarboximide monomer as the second portion of the diblock copolymer for the insulating portion. The latter monomer is synthesized in an identical fashion as the synthesis of oxanorbornenedicarboximide (ref. 28) except phenyl maleimide (TCI-America) was used instead of maleimide. The diblock polymer is synthesized also in the same fashion as the homopolymer except that the second monomer is introduced into the reaction mixture after 30 minutes and is left to react for a similar duration as the first monomer. The reaction is terminated and precipitated also like the homopolymers. The resulting diblock polymer has 50% ethylene oxide monomer by mass compared to the phenyl sidechain monomer. The molecular weights of the homopolymers and the blocks of the diblock were designed to be between 30kDa and 36kDa and dispersities (Đ) are 1.2 or less.

*Sample Preparation*

A 25 mg/mL solution is prepared by dissolving the solid polymer electrolyte in dichloromethane. Unless otherwise noted, Lithium bis(trifluoromethylsulfonimide) (LiTFSI) is added to the solution to achieve a 30% wt/wt LiTFSI to polymer electrolyte ratio. The diblock copolymer system used in imaging is incorporated with 10% wt/wt LiTFSI and otherwise prepared in an identical manner. All solutions are prepared in a nitrogen glovebox to avoid as much water contamination or sample degradation as possible.

*Device Preparation*

Prefabricated glass slides coated with a layer of indium tin oxide are sonicated in solutions of acetone and isopropyl alcohol successively. After sonication, the slides are plasma-cleaned for a period of 10 minutes and then immediately transferred to a dry-nitrogen atmosphere glovebox. Within this glovebox, the slides are spin-coated with the pre-prepared solutions containing the requisite solid polymer electrolyte. This process generates films with an average thickness of ~150 nm. For films to be investigated by EIS, 40 nm gold electrodes are thermally evaporated using a device mask with multiple contacts.

*Atomic Force Microscopy*

AFM measurements were taken on an Asylum Research Cypher with environmental controls in a dry nitrogen environment. A 300 kHz resonance frequency silicon cantilever with a conductive Pt/Cr coating is used. To accomplish the trEFM measurement we use our custom software built within the Igor Pro/Asylum Research environment.

*Device Testing*

Electrical Impedance Spectroscopy is performed using an Autolab PGSTAT302N measurement system with frequency response analyzer. The experiment was performed in a dry nitrogen environment. The opposite contacts are connected to the measurement device which applies a 10 mV potential at a series of decreasing frequencies from 100 kHz to 100 mHz.

### Supporting Information

Along with this work we supply the following supplementary figures:

1. The instrument response to an instantaneous change in applied voltage.
2. A comparison of EIS fitting procedures.

3. An analysis of the spatial resolution of the trEFM technique.